\documentclass[a4paper]{jpconf}
\usepackage{lmodern}        
\usepackage[T1]{fontenc}
\usepackage[utf8]{inputenc} 
\usepackage{graphicx}  
\usepackage[labelformat=simple]{caption}
\usepackage{subcaption}
\usepackage[table,x11names]{xcolor}
\usepackage{amsmath}
\usepackage[pdftex,bookmarksopen,bookmarksnumbered,linktoc=all,hidelinks]{hyperref}

\begin{document}
\title{Constraints of Cosmic Expansion Using an MSF}

\author{Goratamang Gaedie$^{1,2}$, Shambel Sahlu$^{1,3}$, and Amare Abebe$^{1,4}$}

\address{$^1$ Centre for Space Research, North-West University, Potchefstroom 2520, South Africa}
\address{$^2$ South African Astronomical Observatory, Cape Town, 7925, South Africa}
\address{$^{3}$ Department of Physics, Wolkite University, Wolkite, Ethiopia}
\address{$^4$ National Institute for Theoretical and Computational Sciences (NITheCS), South Africa}


\ead{\url{ggaedie@gmail.com}}

\begin{abstract}
In this paper, we propose a modified scale factor (MSF) that allows us to explore the accelerating expansion of the universe without invoking the traditional dark-energy model, as described in the Lambda cold dark matter ($\Lambda$CDM) model. Instead, the MSF model introduces parameters that encapsulate the effects traditionally attributed to dark energy. To test the viability of this MSF, we constrained the model using the observational Hubble parameter (OHD), distance modulus measurements (SNIa), and their combined datasets (OHD + SNIa).
We implement a Monte Carlo Markov Chain (MCMC) simulation to find the best-fit values of the model parameters. The MSF model produced best-fit values for the parameter $p$ associated with the power law of the matter-dominated era and $\beta$, the exponential parameter for the dark-energy-dominated era. For our MSF, these values are $p$ = 0.28 and $\beta$ = 0.52 when using SNIa data, $p$ = 0.63 and $\beta$ = 0.30 for OHD data and $p$ = 0.45 and $\beta$ = 0.53 for a combination of datasets (OHD + SNIa). The numerical results and plots of the deceleration parameter, fractional energy density, Hubble parameter, and luminosity distance are presented which are the key parameters for studying the accelerated expansion of the universe. We compare the results of our model with that of the $\Lambda$CDM model and reconcile them with astronomical observational data. Our results indicate that the MSF model shows promise, demonstrating good compatibility with current astronomical observations and performing comparably to the $\Lambda$CDM model across various datasets, particularly in predicting the accelerating expansion of the universe, while providing a unified framework that incorporates the simultaneous influence of matter and dark energy components.
\end{abstract}

\section{Introduction}
The discovery of accelerating expansion of the universe, first revealed in the late 1990s through observations of distant Type Ia supernovae \cite{riess1998astron, {perlmutter1999astrophys}}, fundamentally altered our understanding of cosmology. This groundbreaking observation led to the introduction of dark energy \cite{freeman2006search, nicolson2007dark}, an enigmatic form of energy that constitutes approximately $70\%$ of the total energy content of the universe and is believed to drive this acceleration. Despite being a crucial component of the universe, the nature of dark energy remains a profound mystery, and its existence poses significant challenges to our current models of cosmology. In recent years, several alternative models to dark energy have been proposed \cite{Smith_2020}, including modifications to general relativity, quintessence models, modifications to the cosmic scale factor, etc. These approaches aim to address shortcomings such as the fine-tuning problem and the cosmological constant problem \cite{Quartin_2008}. The challenge of fine-tuning stems from the vast gap between the energy density of the universe as observed and the theoretical estimates given by quantum field theory, known as the cosmological constant problem \cite{weinberg1989rev}. Fine-tuning in cosmology encompasses two related but distinct issues: the requirement for extremely precise initial conditions, such as the nearly critical density of the universe, and the challenge of explaining parameter values like the remarkably tiny observed cosmological constant, both of which are explored in alternative models like modified gravity (e.g., $f(R)$ gravity, where $R$ is the Ricci scalar), as potential solutions to the shortcomings of the Lambda cold dark matter ($\Lambda$CDM) framework \cite{Tsujikawa2010}. Understanding the accelerating expansion of the universe is crucial for revealing the true nature of dark energy, and its implications stretch beyond cosmology to fundamental physics. If dark energy does not exist, as assumed in modified gravity models that attribute the accelerating expansion to alterations in gravitational dynamics, its nature or implications cannot be directly probed; instead, these models shift the focus to understanding the underlying modifications in gravity that replicate the effects attributed to dark energy. A more accurate model of the universe could reshape our understanding of the energy components that govern cosmic evolution and lead to breakthroughs in particle physics, quantum mechanics, and beyond.

In this study, we propose an MSF model that offers an alternative to $\Lambda$CDM without explicitly invoking the traditional concept of dark energy. Unlike $\Lambda$CDM, which treats matter and dark energy as dominant in different eras, the proposed MSF model unifies a power law with an exponential term to encapsulate the effects traditionally attributed to both matter and late-time accelerated expansion. This approach allows for a unified treatment of cosmic dynamics throughout history, without requiring a separate dark-energy component.

These models can explain the accelerated expansion of the universe, which has the potential to solve the cosmological constant problem. Our MSF is a unified power law function for the matter-dominated era and an exponential function for late-time accelerated expansion. This functional form has been inspired by previous work \cite{aydiner2022late}, which demonstrated its effectiveness in approximating the evolution of the universe through different cosmological eras, providing an alternative explanation for the accelerating universe. Unlike the $\Lambda$CDM model which treats these epochs as distinct and non-overlapping, the MSF model incorporates the coexistence of radiation, matter, and the effects associated with late-time acceleration throughout cosmic history. This unified approach captures the transitions between epochs more naturally while still reflecting the dominant influences during each era.
The MSF model is defined as
\begin{equation}
    \label{eq:5.2}
    a(t) = a_0\left(\frac{t}{t_0}\right)^p e^{\beta\left(\frac{t}{t_0}\right)},
\end{equation}
where $p$ represents the power-law parameter related to the matter-dominated era, $\beta$ is the exponential parameter related to the late-time accelerated expansion, and $t_0$ is the current age of the universe. The term dark-energy-dominated era typically suggests a physical dark-energy component responsible for cosmic acceleration. The MSF model uses an exponential function not tied to a physical dark-energy form but as a mathematical construct to explain the accelerated expansion of the universe. This method eliminates the need for an actual dark-energy field, instead offering a different mechanism to achieve similar results, aligning the terminology with the model's underlying principle. We proposed the physical mechanism of the unified scale factor $a(t)$, MSF that represents two phases of the universe: matter-dominated and dark-energy-dominated eras together to study late-time cosmology.

For the case of the MSF model, the Hubble parameter is expressed as
\begin{equation}
    H(z) = \frac{\beta H_0}{\beta + p}\left(1+ \frac{1}{W\left(\frac{\frac{\beta}{p}e^{\frac{\beta}{p}}}{(1+z)^{\frac{1}{p}}}\right)}\right)\;,
    \label{eq_MSFLCDM}
\end{equation}
where $H_0$ is the Hubble constant, $(1+z)$ describes the scaling of cosmic parameters with redshift, reflecting the universe's expansion and $W$ is the Lambert function, which is the inverse function of $z=We^W$, often used in physics to solve transcendental equations. It helps to express the solution in a compact form that captures the relationship between the redshift and the Hubble parameter in our model.

The Hubble parameter for the $\Lambda$CDM model is expressed as
\begin{equation}
    H = H_0\sqrt{\Omega_{m}(1+z)^3 +1 - \Omega_{m} } ,
    \label{eq_HLCDM}
\end{equation}
where $\Omega_{m}$ is the matter density parameter, $(1+z)^3$ is the factor accounting for the scaling of the matter density with redshift and $1 - \Omega_{m}$ is the dark energy density parameter, $\Omega_{\Lambda}$.
%
%
\section{Results}
\begin{figure}[!ht]
\centering
\includegraphics[width=0.43\textwidth]{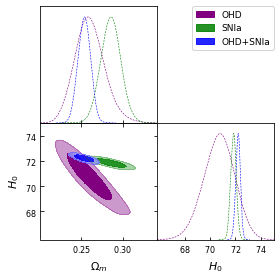}
\includegraphics[width=0.43\textwidth]{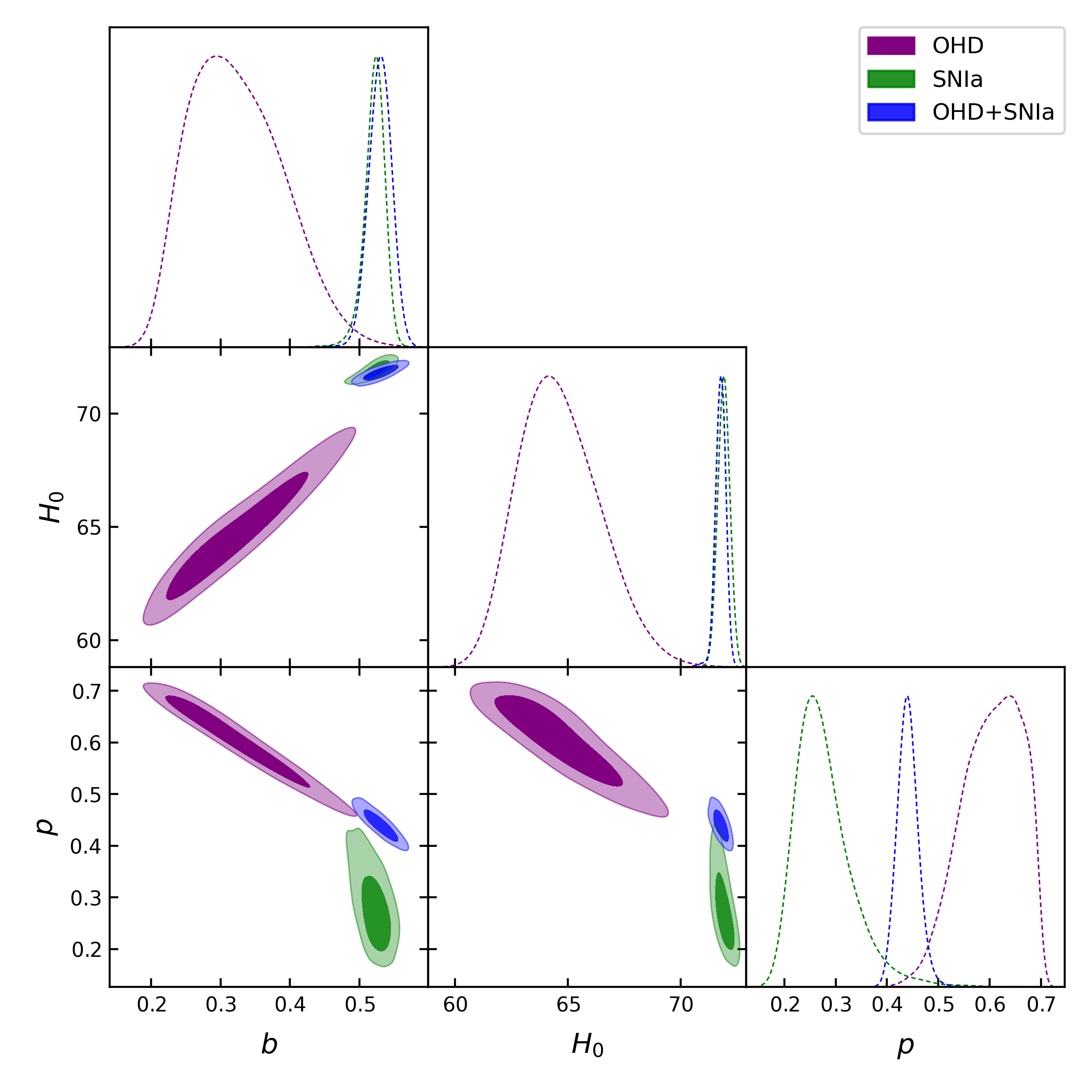}
\caption{\label{fig-sidebyside} (\emph{left}) MCMC simulations for the $\Lambda$CDM model, the Gaussian distribution of the parameters appears smooth and well-defined, indicating stable and consistent parameter estimates with lower uncertainty. (\emph{right}) The MSF model’s parameter constraints are shown using three datasets: observational Hubble data (OHD), Type Ia supernovae (SNIa) data, and the combined OHD+SNIa dataset. The contours represent the 3$\pm\sigma$ confidence region (approximately 99.7\% likelihood), allowing for a direct comparison of parameter distributions across datasets. The broader contours for the MSF model when using OHD alone highlight larger uncertainties, while the combination of OHD and SNIa data results in narrower distributions, reflecting reduced uncertainties due to complementary constraints from the datasets. This behaviour underscores the model’s dependence on dataset selection and the interplay of their respective error margins and calibrations.}
\label{fig:mcmc}
\end{figure}

\begin{figure}[!ht]
\centering
\includegraphics[width=0.43\textwidth]{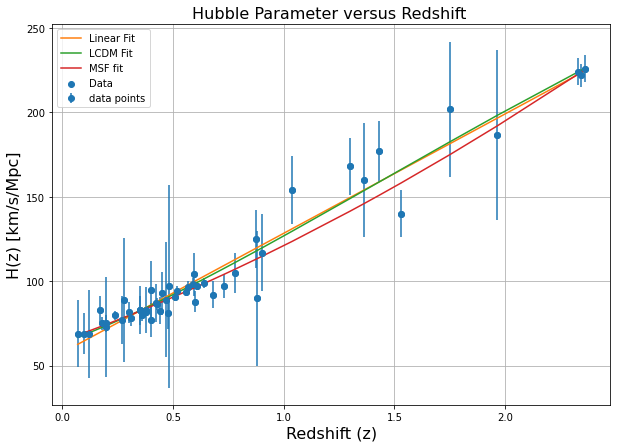}
\includegraphics[width=0.43\textwidth]{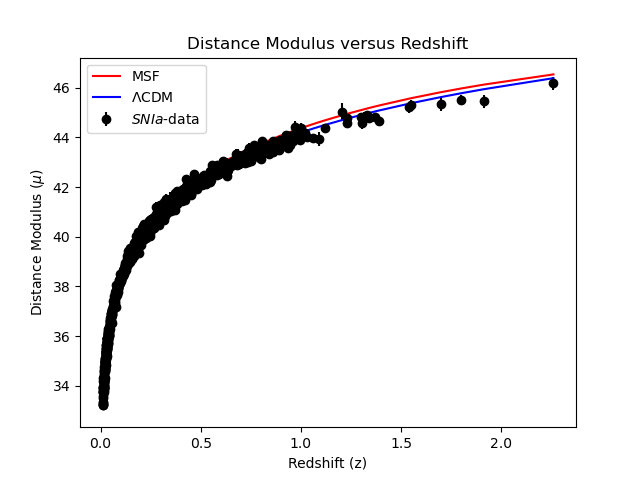}
\caption{\label{fig-HvZ} (\emph{left}) A graph of the Hubble parameter versus redshift is presented, with error bars and a $\Lambda$CDM, MSF (MSF fit) and linear curve-fit lines plotted using OHD data. (\emph{right}) Distance modulus versus redshift graph for the MSF model and the $\Lambda$CDM model plotted with SNIa data.}
\end{figure}

\begin{figure}[ht]
\centering
\includegraphics[width=0.48\textwidth]{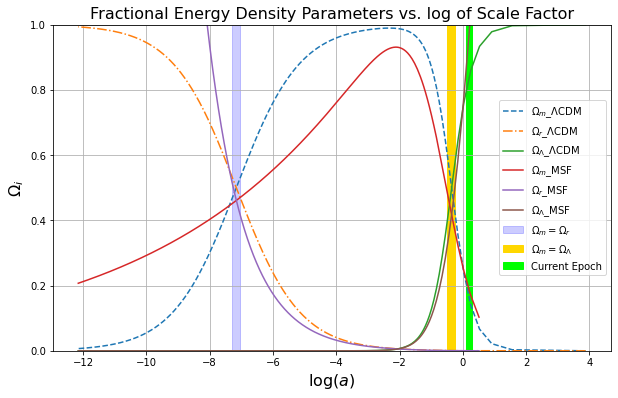}
\includegraphics[width=0.43\textwidth]{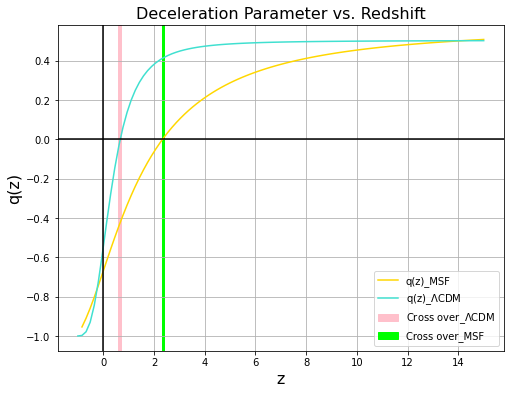}
\caption{ (\emph{left}) Graph of the fractional energy density $\Omega_i$ versus the logarithm of the scale factor for $\Lambda$CDM and MSF, using the best-fit MCMC values from Table \ref{tab:2}. For the $\Lambda$CDM model, the sum $\Omega_i$ is equal to 1, consistent with the assumption of a flat universe in contrast, for the MSF model, the sum of $\Omega_i$ exceeds 1, indicating a deviation from flatness or an additional effective contribution to the energy density. (\emph{right}) Graph plotting the deceleration parameter against the redshift. At a redshift of 0, the value of $q$ is calculated to be -0.55. The universe is decelerating when $q$ is less than 0 and accelerating once $q$ exceeds 0. The redshift is noted to be -0.70 at $q=0$ for the MSF model.}
\label{fig-Q&FED}
\end{figure}

In this section, we present the numerical results, including the contour plots of the parameters in Figure \ref{fig-sidebyside} after constraining the observation parameters through the MCMC simulation. We also present the Hubble diagram in Figure \ref{fig-HvZ} (\emph{left}), the distance modulus in Figure \ref{fig-HvZ} (\emph{right}), the fractional energy density in Figure \ref{fig-Q&FED} (\emph{left}) and the deceleration parameter in Figure \ref{fig-Q&FED} (\emph{right}). The contour plots did not align either at the $1\sigma$ or $2\sigma$ level of confidence. 
 Figure \ref{fig:mcmc} shows the MCMC results for our MSF and the $\Lambda$CDM model for the OHD, SNIa data, and OHD + SNIa data. The Gaussian for $\Lambda$CDM is smooth for both data sets, reflecting a well-constrained parameter space, while the Gaussian distribution for the MSF is less aligned, suggesting greater variability or sensitivity to the data. Specifically, we observe a broader Gaussian distribution when using the OHD data alone, indicating larger uncertainties, and a narrower Gaussian distribution when combining the SNIa and OHD+SNIa data, reflecting reduced uncertainty. This difference indicates that, while the MSF can capture some aspects of cosmic acceleration, its parameter constraints are sensitive to the specific characteristics of each dataset, such as the error bars or calibration issues inherent in SNIa data.

The Gaussian distributions discussed pertain to the range of parameter probabilities determined by MCMC analysis. A broader Gaussian indicates greater uncertainty in parameter estimation, whereas a narrower Gaussian signifies more stringent constraints. This sensitivity might signify the challenges in aligning the MSF with datasets, especially considering discrepancies such as Hubble tension, which is the ongoing variation in $H_0$ values measured using the SNIa dataset. The response of the MSF to the attributes of the dataset might offer a chance to explore or address aspects of the Hubble tension, especially if it uniquely engages with data that are adjusted on varying cosmic scales. The misalignment of the contours in plots such as Figure \ref{fig:mcmc} reflects variances in the parameters estimates between datasets such as OHD and SNIa. This can originate from unique data set sensitivities, including varying error margins and calibration techniques, or from the model's limitations in harmonising the data sets. Additionally, it may signal broader cosmological tensions, like the Hubble tension, where variations arise in the Hubble constant measurements across different cosmic timeframes or scale. These differences emphasise the necessity to explore whether the issue originates from observational biases, model assumptions, or fundamental inconsistencies, pointing towards potential insights into novel physics or model improvements.


\begin{table}[!ht]
    \centering
    \scalebox{1}{
    \begin{tabular}{|c|c|c|c|c|c|c|}
    \hline
        \rowcolor{lightgray} $\Lambda$CDM & $\Omega_{m}$  & $H_0$ & $p$ & $\beta$\\
        \hline
        SNIa &  $0.28^{+0.01}_{\text{--}0.01}$& $71.86^{+1.30}_{\text{--}1.30}$ & - & - \\
        \hline
        OHD & $0.27^{+0.02}_{\text{--}0.02}$ & $70.11^{+1.10}_{\text{--}1.10}$ & - & - \\
        \hline
        OHD+SNIa& $0.25^{+0.07}_{\text{--}+0.07}$ & $72.23^{+0.18}_{\text{--}0.18}$&-&-\tabularnewline
        \hline
       \rowcolor{lightgray} MSF &  & & & \\
        \hline
        SNIa & - &$71.93^{+0.17}_{-0.15}$ & $0.28^{+0.01}_{-0.00}$ & $0.52^{+0.06}_{-0.07}$\\
        \hline
        OHD & - & $64.69^{+2.40}_{-2.38}$ & $0.63^{+0.09}_{-0.08}$ & $0.30^{+0.09}_{-0.10}$\\
        \hline
        OHD+SNIa & - & $71.75^{+0.24}_{-0.20}$ & $0.45^{+0.02}_{-0.03}$ & $0.53^{+0.02}_{-0.01}$ \\
        \hline
    \end{tabular}}
    \caption{The best-fit parameter values for the  $\Lambda$CDM and MSF using MCMC simulation with OHD data, SNIa data, and the combined data sets.}
    \label{tab:2}
\end{table}

\noindent The constraint values represent the optimal parameters obtained through the MCMC analysis, considering $H_0$ and the matter density parameter ($\Omega_{m}$) within the $\Lambda$CDM framework. These parameters were estimated by matching the model to observational data, specifically SNIa and OHD. The close agreement of these values across both datasets highlights the effectiveness of the $\Lambda$CDM model in accounting for cosmic expansion. The presentation of these constraint values underscores the model's robustness and reliability when assessed against different kinds of observational data, serving as a standard for other models. As shown in Table \ref{tab:2}, the constraint values of $\Lambda$CDM for the SNIa and OHD data are nearly identical. In contrast, the MSF model reveals significant differences in constraint values between OHD and SNIa data, indicating possible sensitivities to the type of observational data used.

Figure \ref{fig-HvZ} (\emph{left}) shows that the Hubble parameter increases as the redshift increases. For redshifts below 1 ($z<1$), all models align well with the data points. The $\Lambda$CDM model matches the observed data more closely, especially at redshifts greater than 1 ($z\ge1$). At very high redshifts ($z>2$), every model intersects with the data points. Up to this point, both models fit the data effectively, with the $\Lambda$CDM model fitting better than the MSF model. In Figure \ref{fig-HvZ} (\emph{right}) both the $\Lambda$CDM model and the MSF model aim to fit the data and explain the observed acceleration of the universe. At low redshifts ($z<1$), the two models overlap closely and fit the data equally well. At higher redshifts ($z>1$), the MSF model (red) predicts slightly higher distance modulus values compared to $\Lambda$CDM (blue), although both are still close to the data points. This discrepancy could indicate different treatments of the late-time acceleration between the two models. The good agreement between the data and both models reflects their statistical consistency, as seen in the $\chi^2$ and p values analysed previously. This plot demonstrates that both the MSF and $\Lambda$CDM models effectively describe the observed SNIa data, validating their ability to explain cosmic expansion. The MSF model and $\Lambda$CDM differ in their treatment of late-time cosmic acceleration. $\Lambda$CDM assumes a constant cosmological constant that drives acceleration uniformly across all epochs. In contrast, the MSF model employs a dynamic scalar field whose energy density evolves over time. At low redshifts, the MSF model closely follows the distance modulus predicted by SNIa data, mimicking the behaviour of $\Lambda$. However, at higher redshifts, it deviates slightly, probably due to the evolving dynamics of the scalar field. This indicates that the MSF model allows for a time-dependent influence of dark energy, in contrast to the static nature of $\Lambda$.

Figure \ref{fig-Q&FED} (\emph{left}) shows the fractional energy density $\Omega$, revealing that the MSF model projects a sharper decrease in matter density than the $\Lambda$CDM model at higher values of $a$, underscoring the model's approach to dark-energy and matter coevolution. As demonstrated in Figure \ref{fig-Q&FED} (\emph{right}), at redshifts $z\approx 0$, the MSF model aligns well with the $\Lambda$CDM model. However, the MSF model forecasts a marginally faster acceleration in the universe's expansion at higher redshifts. This is apparent in the deceleration parameter ($q(z)$) graph, where deviations from $\Lambda$CDM predictions become noticeable at $z > 0.7$, indicating a faster expansion during the dark-energy-dominated era. Hence, the MSF model effectively captures the accelerated expansion caused by dark energy, a force opposing gravitational attraction, leading to the universe's increasingly rapid expansion.

\subsection{Statistical Analysis}
To test the performance of the model, we calculated the AIC and BIC for the MSF and $\Lambda$CDM models. The $\Lambda$CDM model, with fewer parameters, performed consistently better on both AIC and BIC metrics, indicating a better fit with less complexity. The MSF model, while providing a good fit, because it introduces more parameters, it loses the width fitting. When you use more parameters, the $\chi^2$ becomes lower for the new parameters. According to statistical selection criteria, if $\Delta IC\leq 2$ there is observational support for the model, for $4\leq \Delta IC \leq 10$ there is less observational support for the model, and when $\Delta IC \geq 10$ there is no observational support for the model \cite{Hough2021ConfrontingTC,{Burnham}}.

\begin{table}[h!]
    \centering
    \scalebox{0.87}{
    \begin{tabular}{|c|c|c|c|c|c|c|c|c|}
    \hline
        \rowcolor{lightgray} $\Lambda$CDM & $\chi^2$ & Reduced $\chi^2$ & Likelihood value & AIC & $\Delta$AIC & BIC & $\Delta$BIC&Fit\\
        \hline
        OHD & 28.53 & 0.59 & -14.26 & 34.53 & - & 40.32 & -& Good fit\\
        \hline
        SNIa &1035.68&0.99&-517.84&1039.68&-&1049.59&-& Good fit\\
        \hline
        OHD+SNIa &1087.41&0.98&-543.70&1091.41&-&1101.43&-&Good fit\\
        \hline
        \rowcolor{lightgray} MSF & $\chi^2$ & Reduced $\chi^2$ & Likelihood value & AIC & $\Delta$AIC & BIC & $\Delta$BIC&Fit\\
        \hline
        OHD & 25.40 & 0.53 & -12.70 & 31.40 & 3.14 & 37.19 & 3.14& Good fit\\
        \hline
        SNIa &1033.94&0.99&-516.97&1039.94&0.26&1054.80&5.21&Good fit\\
        \hline
        OHD+SNIa &1092.06&1.00&-546.03&1098.06&6.65&1113.06&11.63&Good fit\\
        \hline
    \end{tabular}}
    \caption{The best-fit statistical values of $\chi^2$ obtained via MCMC simulation for OHD data, SNIa data, and their combination. The likelihood value is log(L).}
    \label{tab:3}
\end{table}
Together, these statistical tools provide a comprehensive view of the performance of the model. Although the $\Lambda$CDM model is preferred due to its simplicity, the complexity of the MSF model allows greater flexibility to fit the data, particularly for OHD measurements. Balancing this flexibility with the additional degrees of freedom is essential when evaluating the viability of alternative cosmological models.
By setting the values of $\beta = 0.301$, $p= 0.633$,  we obtain the age of the universe $t_0 = 13.7$ Gyrs, which is consistent with the observationally determined age of the universe from cosmological probes, such as the results of the Planck satellite $t_0 = 13.8$ Gyr. These parameters are significant because they not only produce a realistic age of the universe but also suggest that the MSF model can achieve results close to those of the widely accepted $\Lambda$CDM.

\section{Discussions and Conclusion} 

Our evaluation indicates that the modified scale factor model (MSF) aligns well with the OHD data. In particular, it achieves a $\chi^2$ value marginally below that of the $\Lambda$CDM model, highlighting its robustness in fitting these data. However, the MSF model struggles to fit the SNIa data independently as effectively as the $\Lambda$ CDM model, although the difference is marginal. For the combined OHD+SNIa dataset, the MSF model still performs well, but shows weaker support when assessed using information criteria such as AIC and BIC. This suggests that while the MSF model is a feasible alternative to $\Lambda$CDM, it requires further refinement and broader validation to address inconsistencies in the datasets, especially the SNIa data.

Specifically, the statistical results show:
\begin{itemize}
    \item For OHD data at $p = 0.63$, $\beta = 0.30$ and $H_0 = 71.93$, the MSF model achieves a minimum value $\chi^2$ of 25.40 with a reduced $\chi^2$ of 0.53, slightly outperforming $\Lambda$CDM ($\chi^2$ = 28.53, reduced $\chi^2$ = 0.59). Although this might indicate a strong fit, such low reduced values of $\chi^2$ may suggest overfitting or that the residuals deviate from Gaussian assumptions. This could reflect underestimated observational errors or unaccounted for systematic effects in the data. With $\Delta$AIC = 3.14 and $\Delta$BIC = 3.14, the MSF model is moderately supported compared to the $\Lambda$CDM.
    \item For SNIa data at $p = 0.28$, $\beta = 0.0.52$ and $H_0 = 64.69$, the MSF model achieves a minimum $\chi^2$ value of 1033.94, a reduced $\chi^2$ of 0.99, and values of $\Delta$AIC and $\Delta$BIC of 0.26 and 5.21, respectively. This reflects adequate but not strong support for the MSF model compared to $\Lambda$CDM.
    \item For the combined OHD+SNIa dataset at $p = 0.45$, $\beta = 0.53$ and $H_0 = 71.75$, the MSF model produces a minimum $\chi^2$ value of 1092.06 with a reduced $\chi^2$ of 1.00, slightly higher than $\Lambda$CDM ($\chi^2$ = 1087.41, reduced$\chi^2$ = 0.98). The $\Delta$AIC and $\Delta$BIC values of 6.65 and 11.63 indicate weak observational support for the MSF model in this scenario.
    \item Our investigation reveals that the MSF and $\Lambda$CDM models both align well with the OHD data, showing probability values (p-values) ($P(\ge \chi^2)$) close to 1 (0.998 and 0.991, respectively), which means that the observations $\chi^2$ correspond well to the models. In contrast, for the SNIa data, the p-values decrease to approximately 0.6 for both models, suggesting that neither model fits these data perfectly. Although the MSF model demonstrates slightly better performance than $\Lambda$CDM for OHD data, it offers only marginal support for the combined OHD+SNIa dataset, with a p-value of 0.53, while the $\Lambda$CDM model presents a p-value of 0.57 for the same data set. These findings imply that, while the MSF model has potential as a viable alternative to $\Lambda$CDM, further adjustments are needed, particularly to better fit the SNIa data.

\end{itemize}
The values of $\Delta$AIC and $\Delta$BIC support the notion that the $\Lambda$CDM model still fits the data more effectively overall, particularly for the combined data set. Nevertheless, the performance of the MSF model, especially for the OHD data, highlights its potential as an alternative cosmological model.
In general, this model needs further investigation in future research work against more precise datasets, such as baryon acoustic oscillations (BAO) and cosmic microwave background (CMB) observations, to further evaluate its viability.  One limitation of our study is the assumption of homogeneity and isotropy in the universe, which might overlook small-scale anisotropies. Furthermore, the different parameter estimates across various datasets suggest potential systematic biases, as well as non-colocation of inferred parameters, which may arise due to inconsistencies between data types or model assumptions. Addressing these discrepancies will be crucial in future research to establish the robustness of the MSF model.

\section*{References}
\bibliographystyle{321-Gaedie}
\bibliography{321-Gaedie}

\providecommand{\newblock}{}
\begin{thebibliography}{10}
\expandafter\ifx\csname url\endcsname\relax
  \def\url#1{{\tt #1}}\fi
\expandafter\ifx\csname urlprefix\endcsname\relax\def\urlprefix{URL }\fi
\providecommand{\eprint}[2][]{\url{#2}}

\bibitem{riess1998astron}
Riess A~G {\em et~al.\/} 1998 {\em The Astronomical Journal\/} {\bf 116} 1009

\bibitem{perlmutter1999astrophys}
Perlmutter S, Collaboration S~C~P {\em et~al.\/} 1999 {\em Astron. J\/} {\bf 116} 1009

\bibitem{freeman2006search}
Freeman K and McNamara G 2006 {\em In search of dark matter\/} (Springer)

\bibitem{nicolson2007dark}
Nicolson I 2007 {\em Dark Side of the Universe\/} (Johns Hopkins University Press)

\bibitem{Smith_2020}
Smith M {\em et~al.\/} 2020 {\em The Astronomical Journal\/} {\bf 160} 267

\bibitem{Quartin_2008}
Quartin M, Calvão M~O, Jorás S~E, Reis R~R~R and Waga I 2008 {\em Journal of Cosmology and Astroparticle Physics\/} {\bf 2008} 007

\bibitem{weinberg1989rev}
Weinberg S 1989 {\em Rev. Mod. Phys\/} {\bf 61}

\bibitem{Tsujikawa2010}
Tsujikawa S 2010 {\em Modified Gravity Models of Dark Energy\/} (Springer Berlin Heidelberg) pp 99--145

\bibitem{aydiner2022late}
Aydiner E, Basaran-{\"O}z I, Dereli T and Sarisaman M 2022 {\em The European Physical Journal C\/} {\bf 82} 39

\bibitem{Hough2021ConfrontingTC}
Hough R~T, Sahlu S, Sami H, Elmardi M, Swart A and Abebe A 2021 {\em International Journal of Modern Physics D\/}

\bibitem{Burnham}
{Burnham} K~P and {Anderson} D~R 2004 {\em Sociological Methods and Research\/} {\bf 33} 261--304

\end{thebibliography}

\end{document}